\documentclass[aps,prd,superscriptaddress,showpacs,nofootinbib]{revtex4}
\usepackage{graphicx}
\usepackage{epstopdf}
\usepackage{amsmath}
\usepackage{amsfonts}
\usepackage{amssymb}
\usepackage{latexsym}

\begin{document}
\title{Chern-Simons-Schwinger model of confinement in $QCD$}
\author{Antonio Aurilia}
\email{aaurilia@cpp.edu}
\affiliation{Department of Physics and Astronomy, California State Polytechnic University-Pomona, Pomona, California 91768, USA}
\author{Patricio Gaete}
\email{patricio.gaete@usm.cl}
\affiliation{Departmento de F\'{\i}sica and Centro Cient\'{\i}fico-Tecnol\'ogico de Valpara\'{\i}so, Universidad T\'{e}cnica Federico Santa Mar\'{\i}a, Valpara\'{\i}so, Chile}
\author{Euro Spallucci}
\email{spallucci@ts.infn.it}
\affiliation{Dipartimento di Fisica Teorica, Universit\`a di Trieste and INFN, Sezione di Trieste, Italy}
\date{\today}

\begin{abstract}
It has been shown that the mechanism of formation of glue-bags in the strong coupling limit of Yang-Mills theory can be understood in terms of the dynamics of a higher-rank abelian gauge field, namely, the 3-form dual to the Chern-Simons topological current. 

Building on this result, we show that the field theoretical interpretation of the Chern-Simons term, as opposed to its topological interpretation, also leads to the analytic form of the confinement potential that arises in the large distance limit of $QCD$. In fact, for a $(3+1)$-dimensional generalization of the Schwinger model, we explicitly compute the interaction energy. This generalization is due to the presence of the topological gauge field $A_{\mu\nu\rho}$. Our results show that the static potential profile contains a linear term leading to the confinement of static probe charges.

Once the quantum effects of the axial vector anomaly in $QCD$ are taken into account, the new gauge field and its matter-current counterpart provide an exact replica of the Schwinger mechanism of ``charge-screening'' that operates in $QED_2$. 
\end{abstract}
 \maketitle

\section{Introduction and synopsis}

An analytical proof of color confinement in $QCD$ still eludes us in spite of many ingenious model calculations and definite hints from computer simulations based on lattice gauge-theory. Computer simulations also provide some evidence for the existence of glueballs as bound states of pure glue that cannot be accounted for by 
conventional perturbative techniques.

The root of the problem is well known: while asymptotic freedom is a well established property of the perturbative dynamics of $QCD$, the transition to infrared slavery is problematic because of non-perturbative effects that dominate in the large distance limit of the theory. Once this "large distance limit" is defined in terms of some phenomenological scale of distance, the immediate problem is that of identifying the dynamical variables that operate in that limit. A hint about the nature of those hidden dynamical variables comes from the phenomenological bag models of hadrons \cite{Hasenfratz:1977dt}: the partial success of those models indicate that, in the large distance limit of $QCD$, the \textit{spatial extension} of hadrons and the bag degrees of freedom must somehow be included among those new dynamical variables. On the other hand, in order to speak meaningfully of a "$QCD$-solution" of the confinement problem, one would expect that \textit{such variables should arise from the very dynamics of $QCD$} and control the mechanisms of color confinement and hadronization through charge screening. 

A significant step that meets the above expectation was taken by Gabadadze several years ago \cite{Gabadadze:1997kj} building on the early suggestion that \textit{there is a hidden long range force} \cite{Luscher:1978rn} 
\textit{in the topological sector of YM-theory defined by the so called $\theta$-term} \cite{Jackiw:1976pf,Callan:1976je}. The prevalent ``instanton interpretation'' of that term \cite{Belavin:1975fg} has prevented an earlier and wider recognition of that long range force and its relevance not only for the formation of glueball states but also for the fundamental problem of confinement and screening of color charges.

The focus on instanton solutions in YM-theory renders the confinement problem cumbersome, if not intractable: on the one hand, instanton-configurations of non-abelian gauge fields play an essential role in determining the structure of the Yang-Mills theory vacuum and are responsible for tunneling processes among topologically distinct vacuums. On the other hand, they fail to satisfy the widely accepted criterion for color confinement, namely, the so called "area law" for the Wilson loop \cite{Wilson:1974sk}. As noted in  \cite{Gabadadze:1997kj}, this failure suggests that the formation of non-perturbative bound states, such as glueballs, requires at least a strong correlation among instantons. However, the study of instanton effects in the strong coupling limit of YM-theory is a cumbersome task precisely because of the lack of non-perturbative computational techniques.
A notable advance was made several years ago through the string theory spin-off, now universally known as $AdS/CFT$, or $AdS/QCD$ 
\cite{Maldacena:1997re}. For an authoritative review see \cite{Aharony:1999ti}. Capitalizing on the full power of the string theory machinery, this new approach offers effective non-perturbative techniques to deal with confinement in super-symmetric Yang-Mills theory in the Large-$N$ limit \cite{Witten:1998zw}. However, we believe that a satisfactory proof of confinement in real $QCD$ with $N=3$, where both  super-symmetry and scale invariance are broken, is still missing and this is the rationale for considering an alternative approach grounded in the more traditional framework of quantum field theory. 

In the following, we shall argue that the key to the resolution of this quandary is to shift from the conventional topological interpretation of the Chern-Simons term to a field theoretical interpretation based on the hidden long range force as originally suggested in Refs.\cite{Luscher:1978rn,Aurilia:1979dw}. Indeed, this was the basic recognition made by Gabadadze who successfully applied the field theoretical approach 
to the problem of formation of non-perturbative glueball states 
\cite{Gabadadze:1997zc}. 

On our part, we will show that the study of that hidden long range force in $QCD$ casts the problem of color confinement in a completely new perspective, one in which calculations can be carried out \textit{exactly} and whose physical interpretation is transparent.

The implementation of this idea involves three distinct steps that will be discussed in the following sections:
first, we will give a precise meaning to the ``field theoretical interpretation'' of the Chern-Simons term in $QCD$ and define the ``long range force'' that arises from that interpretation; second, we shall derive the form of the static potential associated with that force and show that it satisfies the Wilson loop criterion for confinement; third, we will evaluate the quantum effects of the axial anomaly in QCD on the static potential in order to account for the phenomenon of ``~color charge screening~''. 

The guiding principle of the above research plan is the correspondence, noted long ago in Ref.\cite{Aurilia:1979dw} but never fully exploited, between the colorless topological sector of QCD and the zero-charge sector of QED in (1+1)-spacetime dimensions. The latter is also known as the ``Schwinger model'' \cite{Schwinger:1962tn,Schwinger:1962tp} and the broader objective of this paper is to give a precise meaning to the above correspondence. This extrapolation from two to four dimensions is important because the occurrence of a rising Coulomb potential, the ``screening'' of it through the generation of mass and the subtle interplay with gauge invariance represent the essential ingredients 
of a mechanism by which one hopes to understand the phenomenon of quark-binding into physical hadrons. These issues were first analyzed in $QED_2$ in \cite{Schwinger:1962tn,Schwinger:1962tp,Casher:1973uf,Coleman:1975pw}.

On the mathematical side, it seems useful, even at this introductory stage, to point to the nature of the long range force that is inherent, but not manifestly present, in the topological formulation of the Chern-Simons term in $QCD$. In a nutshell, this force is mediated by a higher rank abelian gauge potential $A_{\mu\nu\rho}$ which is dynamically realized, in $QCD$, as the well-known composite, colorless, combination of YM potential  $\mathbf{A}_\mu $ 
and field strength $\mathbf{F}_{\nu\rho}$, namely, \textit{the dual of the Chern-Simons current}
\begin{eqnarray}
A_{\mu\nu\rho} &&\equiv  \frac{1}{\Lambda_{QCD}^2}\,
Tr\,\left(\, \mathbf{A}_{[\, \mu}\,
\mathbf{F}_{\nu\rho\,]}\, \right)
=\frac{1}{\Lambda_{QCD}^2}\,
Tr\left(\, \mathbf{A}_{[\, \mu}\, \partial_\nu\,
\mathbf{A}_{\rho\,]} +\frac{2g}{3} \mathbf{A}_{[\, \mu}\, \mathbf{A}_\nu\,
\mathbf{A}_{\rho\,]}\,\right)\nonumber\\
&&=\frac{1}{\Lambda_{QCD}^2}\,
\left(\, \delta_{ab}\, A^a{}_{\, [\, \mu}\, \partial_\nu\,
A^b_{\rho\,]} +\frac{2g}{3}\, f_{abc}\, A^a_{[\, \mu}\, A^b_\nu\,
A^c_{\rho \, ]}\,\right)\ .\label{topfield}
\end{eqnarray}

In the above expression we have identified $\Lambda_{QCD}$ as the energy scale at which the dynamics of $QCD$ cannot be treated perturbatively. The introduction of this distance scale, at this stage, is required in order to give canonical dimensions to $A_{\mu\nu\rho}$. The confining nature of this higher rank potential was anticipated long ago in Refs.\cite{Aurilia:1978qs,Aurilia:1979dw}  and the manifold  applications of the 3-index potential in particle physics and cosmology have been investigated extensively in Refs.\cite{Aurilia:1977jz,Aurilia:1978qs,Aurilia:1979dx,Aurilia:1978yc}, \cite{Aurilia:1980jz,Aurilia:1980xj}, \cite{Aurilia:1983ih,Aurilia:1984cm,Aurilia:1987cp,Aurilia:1989sb}, \cite{Ansoldi:1997hz,Ansoldi:2000ge,Ansoldi:2001xi,Aurilia:1981xg}, \cite{Aurilia:2004cb,Aurilia:2004fz}. Presently, since the implementation of the Schwinger mechanism in $QCD$ 
depends critically on the properties of the mathematical construct (\ref{topfield}), we shall refer to it as the \textit{Chern-Simons-Schwinger potential,} or, CSS-potential for short.

The next obvious question that arises in the field-theory interpretation of the Chern-Simons term concerns the nature of the ``sources'' of the force mediated by $A_{\mu\nu\rho}$. Here is where we depart substantially from the conventional interpretation of the topological charge density. In the next section we show that the CSS-potential (\ref{topfield}) is subject to a gauge transformation that dictates the nature of its coupling to matter just as in the familiar case of ordinary electrodynamics. However, unlike the vector electromagnetic potential, the 3-vector $A_{\mu\nu\rho}$ does not couple to point-like objects; rather, \textit{it couples to relativistic closed membranes} which, in our interpretation, constitute the physical boundary of hadronic ``bags'' \cite{Aurilia:1979dw}.

Apart from the noteworthy results obtained by Gabadadze in the study of glue-bags, the physical payoff of the new formulation can be summarized thus: first, unlike the Euclidean topological interpretation of the Chern-Simons current, the dynamics of the $3$-vector gauge field (\ref{topfield}) is \textit{exactly solvable in four-dimensions} as is the dynamics of the corresponding $1$-vector gauge field in \textit{two dimensions.} Indeed, solving this field equations for the CSS-potential, one finds that the gauge field $A_{\mu\nu\rho}$ does not transmit physical quanta but gives rise to the same confining potential that operates in classical electrodynamics in ($1+1$)- dimensions. Equivalently, the binding potential between two surface elements on the boundary of a bag is shown to satisfy the Wilson loop criterion for confinement. Furthermore, at the quantum level a mechanism sets in, identical to the "Schwinger mechanism" of $QED_2$, leading to the screening of color charges and "hadronization" in the form of massive pseudo-scalar particles as a consequence of the chiral anomaly in $QCD$. It seems noteworthy that this mechanism of mass generation is, a) fundamentally different from the conventional Higgs mechanism which is based on spontaneous symmetry breaking, and b) it is also a \textit{universal} mechanism in the sense that the dynamics of the CSS-potential can be formulated in any number of spacetime dimensions and may have interesting cosmological consequences. Finally, as announced earlier, the variables that control the dynamics of $QCD$ in the large distance limit are clearly identified. They are, i) the current associated with a relativistic (closed) membrane as the "matter" constituent of the action, and ii) a Maxwell field of the "fourth kind" as the field constituent of the action.

These points are discussed in a self-contained manner in the following sections. We begin, in Section II, with an overview of the reinterpretation of the topological term in YM-theory as it constitutes the basis of our approach. In Section III, for a $(3+1)$-dimensional generalization of the Schwinger model, we explicitly compute the interaction energy between external probe charges. As a result, we obtain a linear term leading to the confinement of static probe charges. As expected, the above potential profile is analogous to that encountered in the usual Schwinger model. Finally, in Section IV, we cast our Final Remarks. Next, in Appendix, we introduce a new way to obtain the anomaly induced effective action with emphasis on the screening effect and mass generation mechanism by way of the quantum anomaly in QCD.

\section{The case for an abelian gauge field of the fourth kind in $QCD$}
\subsection{A new look at the topological charge density in YM-theory}

Consider the so called $\theta$-term in the total action for YM-theory,
\begin{eqnarray}
S_\theta &&\equiv \frac{\theta}{32\pi^2}\,\int d^4x\,
  \, Tr\left(\,\mathbf{F}_{\mu\nu}^\ast\, \mathbf{F}^{\mu\nu}\,\right)\ .
\end{eqnarray}

The $\theta $-dependence of the vacuum energy density in the large $N$-limit has been studied extensively by various methods, even in terms of a $D$-brane construction of YM-theory within the context of string theory \cite{Ansoldi:2000ge}. However, the essential point of our discussion is that the above term admits a field theory interpretation in terms of a \textit{Lagrangian system} from which one can extract some physical properties that are far from evident in the conventional topological interpretation.

The key idea is to reformulate $S_\theta$ in terms of the CSS-gauge potential anticipated in the Introduction. This is possible because of the following well known identities in YM-theory: the integrand of the ${\theta}$-term, namely, "the topological charge density" $Q$ is identically equal to the divergence of the Chern-Simons current 
$K_\mu $ so that $Q=\partial_\mu\, K^\mu = {1\over 32\pi^2}\,\mathrm{Tr}\,{\bf F}_{\mu\nu}\, {\tilde {\bf F}}^{\mu\nu}\equiv \bf{F}{\tilde {\bf F}}$. On the other hand, the Chern-Simons current is the Hodge dual of $A_{\mu\nu\rho}$, $~K^{\mu}={1\over 3!}\,\varepsilon^{\mu\nu\alpha\beta}\, A_{\nu\alpha\beta}~$.
Putting together the above identities, one arrives at the following expression for $Q$,
\begin{eqnarray}
Q = {1\over 4!}\,\varepsilon_{\mu\nu\rho\sigma}\, F^{\mu\nu\rho\sigma}, 
\nonumber
\end{eqnarray}
where the differential four-form
\begin{eqnarray}
F\equiv {1\over 4!}\, F_{\mu\nu\rho\sigma}\, dx^\mu
\wedge dx^\nu \wedge dx^\rho
\wedge dx^\sigma, \nonumber
\end{eqnarray}
represents the field strength for the three-form potential $F=dA$, where
$A~\equiv~{1\over 3!}A_{\nu\rho\sigma}\, dx^\nu \wedge dx^\rho \wedge dx^\sigma $.

We shall refer to $F$ as a ``Maxwell field of the fourth kind'' because of its invariance under a generalized gauge transformation to be specified shortly.

As we shall discuss in the following, the action $S_\theta$ in YM-theory represents the coupling of the $F$-field to 
the bulk of a finite vacuum domain, or "bag" \cite{Aurilia:1978qs}, \cite{Aurilia:1977jz}
\begin{eqnarray}
S_\theta=\theta \int_V Q\, d^4x =-{\theta \over 4!}\,\int_V
F_{\mu\nu\rho\sigma}\, dx^\mu \wedge dx^\nu \wedge dx^\rho
\wedge dx^\sigma \equiv -\theta \int_V F \ , \nonumber
\end{eqnarray}
where $V$ is the region where the topological charge is different from zero.More instructively, using Stokes' theorem, the same action takes the form
\begin{eqnarray}
S_\theta=-\theta \int_V F=-\theta \int_{\partial V} A=
-{\theta \over 3!} \int_{\partial V} A_{\nu\rho\sigma}~
dx^\nu \wedge dx^\rho \wedge dx^\sigma,
\end{eqnarray}
in which the role of the ${\theta}$-angle is now that of a "coupling constant" between the CSS-potential and the conserved 3-form current $J$ associated with the closed membrane that constitutes the physical boundary of the bag:
\begin{equation}
S_\theta =\frac{1}{3!} \, \frac{3\theta}{4\pi^2}\, \int d^4x
\, J^{\lambda\mu\nu}_{\partial V}\,
\mathrm{Tr}\left(\,\mathbf{A}_{[\,\lambda}\,\mathbf{D}_\mu
  \, \mathbf{A}_{\nu\,]}\,\right),
\label{sint}
\end{equation}

\begin{equation}
 J^{\nu\rho\lambda}_{\partial V}\left(\, x\,\right) =
 \int_{\partial V}  \delta^{4)}\left[\, x  - y\,\right] dy^\nu\wedge
dy^\rho\wedge
dy^\lambda, \label{curr}
\end{equation}

\begin{equation}
\partial_\mu J^{\mu\nu\rho}_{\partial V}=0.
\end{equation}

As anticipated in the Introduction, this deceptively simple procedure clearly identifies the variables that govern the formation and dynamics of glue bags. \textit{To the extent that such variables arise through a reformulation of the "topological charge density'', they are an integral part of the dynamics of YM-theory.}
The form of the full Lagrangian for gluodynamics includes a kinetic term for the field strength $F$ introduced above \footnote{This form of the Lagrangian is reported in Ref.\cite{Aurilia:1977jz} as Eq.(5). The mathematical and physical properties of this  Lagrangian system have been studied extensively \cite{Aurilia:1979dx,Aurilia:1978yc,Aurilia:1981xg}, not  only in connection with the dynamics of hadronic bags \cite{Aurilia:1978qs,Aurilia:1984cm,Aurilia:1987cp}  but also in connection with a production mechanism of dark matter \cite{Ansoldi:2001xi} and the cosmological inflationary scenario of the early universe \cite{Aurilia:1983ih,Ansoldi:1997hz}. To be precise, the complete Lagrangian also includes a kinetic term for the closed membrane with or without the presence of gravity. The most general case was discussed in Ref.\cite{Aurilia:1978qs}. Clearly, in the case of $QCD$ the above Lagrangian is gravity independent and the coupling of the CSS-potential is to external membranes nucleated out of the vacuum energy background provided by the $F$-field, as discussed in the text.}
\begin{equation}
 S_A^0\equiv -\frac{1}{2\times 4!}\,\int d^4x \, \left(\, \partial_{[\,
 \mu}\, A_{\nu\rho\sigma\,]}\,\right)^2\ .
\end{equation}

The remarkable properties of the above Lagrangian will be highlighted 
in the following subsection.

\subsection{The meaning of the CSS-Lagrangian}

The computation of the quantum vacuum pressure that determines the structure of the ground state of strong interactions must take into account the contribution of zero-point oscillations of a rank-three gauge field, $A_{\mu\nu\rho}$ \cite{Aurilia:2004cb}. As we discuss below, this field is known to have no radiative degrees of freedom in four dimensions, the only dynamical effect being a static long-range force within finite vacuum domains, or glue bags. The overall effect of this force is a constant but otherwise arbitrary pressure within the bags. This remarkable feature was exploited to associate the $A_{\mu\nu\rho}$-field with the ``bag constant'' of the hadronic vacuum as a kind of \textit{Casimir effect for strong interactions} \cite{Aurilia:2004fz,Aurilia:2004cb}. In other words, the gauge field $A_{\mu\nu\rho}$ does not correspond to a massless particle but gives an energy proportional to the volume when 
external bags are minimally coupled to it \footnote{With hindsight, this work is complementary to that of Ref.\cite{Gabadadze:1997zc} which focuses, instead, on the volume dependence of the topological susceptibility in YM-theory.}.

The gauge potential $A_{\mu\nu\rho}$ gives rise to a "Maxwell field of the fourth kind" in the sense that the corresponding field strength $F_{\mu\nu\rho\sigma}\equiv\partial_{[\,\mu}\, A_{\nu\rho\sigma\,]}$  satisfies the generalized Maxwell equation
\begin{equation}
\partial_\mu\, F^{\mu\nu\rho\sigma}=0\label{max},
\end{equation}
and is invariant under the extended gauge transformation
\begin{equation}
\delta A_{\mu\nu\rho}=\partial_{[\,\mu}\,\lambda_{\nu\rho\,]} \ .
\label{gauge}
\end{equation}

The ``gluonic'' gauge potential, or CSS-potential (\ref{topfield}), is subject to the same extended gauge transformation (\ref{gauge}) in the sense that, when the YM-potential is gauge rotated
\begin{eqnarray}
&& \mathbf{U}\equiv e^{i \mathbf{\Lambda}(x) }=e^{i
\Lambda_a(x)\,\mathbf{T}^a}\ ,\qquad   \mathbf{\Lambda}(x)\equiv
            \Lambda_a(x)\,\mathbf{T}^a,       \\
&& \mathbf{A}_\mu{}^\prime= \mathbf{U}^{-1}\, \mathbf{A}_\mu\,\mathbf{U}-
\frac{i}{g}\, \mathbf{U}^{-1}\,\mathbf{D}_\mu\,\mathbf{U},\\
&& A_{\mu\nu\rho}^\prime=Tr\left(\, \mathbf{A}_{[\, \mu}\,
\mathbf{F}_{\nu\rho\,]}\, \right)^\prime,
\end{eqnarray}
it transforms as follows
\begin{equation}
A_{\mu\nu\rho}^\prime =A_{\mu\nu\rho}-\frac{i}{g}\, Tr\left[\,
\left(\,\mathbf{D}_{[\,\mu}\,\mathbf{U}\,\right)\,
\mathbf{U}^{-1}\,\mathbf{F}_{\nu\rho\,]}\,\right],
\end{equation}
\begin{equation}
\mathbf{D}_\mu \,\mathbf{U}= i\, \mathbf{D}_\mu
\,\mathbf{\Lambda}\,\mathbf{U},
\end{equation}
\begin{eqnarray}
\delta A_{\mu\nu\rho}&& =\frac{1}{g}\,
Tr\left[\,\left(\,\mathbf{D}_{[\,\mu}\,\mathbf{\Lambda}\,\right)\,
\mathbf{F}_{\nu\rho\,]}\,\right]
\equiv \frac{1}{g}\,\partial_{[\,\mu}\,\Lambda_{\nu\rho\,]}.
\label{extgt}
\end{eqnarray}

The gauge transformation (\ref{gauge}) is instrumental in determining the dynamics of the CSS-potential since it \textit{requires} the coupling of the potential to the rank-three current density $J^{\mu\nu\rho}\left(\, x\,\right)$ that we have anticipated in the previous subsection. Thus, it seems noteworthy that, unlike the \textit{phenomenological} bag models of hadrons, the spatial extension of glueballs is a dynamical consequence of the hidden gauge 
invariance of the Chern-Simons term in $QCD$.

Summing up our discussion so far, the full dynamics of $A_{\mu\nu\rho}(x)$, in the presence of external membranes, is 
governed by the Lagrangian density
\begin{eqnarray}
L&&= -{1\over 2\cdot 4!}\,\left(\, \partial_{[\,\mu}\, 
A_{\nu\rho\sigma\,]}\,\right)^2
-{\kappa\over 3!}\, J^{\mu\nu\rho} \,  A_{\mu\nu\rho}  \nonumber\\
&&=-{1\over 2\cdot 4!}\, F^{\lambda\mu\nu\rho}\, F_{\lambda\mu\nu\rho}
+{1\over  4!}\, F^{\lambda\mu\nu\rho}\,\partial_{[\,\lambda}\,
  A_{\mu\nu\rho\,]}-{\kappa\over 3!}\, J_{\partial V}^{\mu\nu\rho} \,  
  A_{\mu\nu\rho}.
  \label{med}
\end{eqnarray}
The remarkable property of the system (\ref{med},\ref{curr}) is that 
its dynamics is \textit{exactly solvable.}

In the absence of coupling the solution is immediate since Maxwell's equation (\ref{max}) is identically satisfied with $F_{\mu\nu\rho\sigma}=f\,\epsilon_{\mu\nu\rho\sigma}$.  The arbitrary integration constant $f$ is physically associated with the vacuum energy density \cite{Aurilia:1978qs,Aurilia:1980xj} \footnote{It has been shown that in higher dimensional models this constant is "\textit{quantized}". This result could provide a solution to the long-standing cosmological constant puzzle \cite{Bousso:2000xa,Dvali:2004tma}.} through the energy momentum tensor
\begin{equation}
 T_{\mu\nu}=\frac{1}{3!}
 F_{\mu\alpha\beta\gamma}\, F_{\nu}^{\alpha\beta\gamma}-
\frac{1}{2\cdot 4!}\,\delta_{\mu\nu}\, F^{\alpha\beta\gamma\delta}\,
F_{\alpha\beta\gamma\delta},
\end{equation}
which, in view of the given expression of the field strength, reduces to the  simple form
\begin{equation}
T_{\mu\nu}=\frac{f^2}{2}\,\delta_{\mu\nu}\ .\label{tensor}
\end{equation}

When the coupling to external bags is switched on, as in (\ref{med},\ref{curr}), the dynamics is still exactly solvable and one way to obtain the general solution is through the following steps. First, the field equations
\begin{eqnarray}
&& \frac{\delta L }{\delta F^{\lambda\mu\nu\rho}}=0\longrightarrow
F_{\lambda\mu\nu\rho}=\partial_{[\,\lambda}\, A_{\mu\nu\rho\,]},\\
&& \frac{\delta L }{\delta A^{\mu\nu\rho}}=0\longrightarrow
\partial_\lambda \, F^{\lambda\mu\nu\rho}= \kappa\, J^{\mu\nu\rho} (x)\ ,
\label{maxx}
\end{eqnarray}
imply that the current is conserved, a property that is consistent with the  invariance of the system under the extended gauge transformation (\ref{gauge}):
\begin{equation}
\delta A_{\mu\nu\rho}=\partial_{[\,\mu}\,\lambda_{\nu\rho\,]}
\longleftrightarrow
\partial_\mu\, J^{\mu\nu\rho} (x)=0\ .  \label{inc}
\end{equation}

In this instance, the conserved membrane current $J$ can be written as the divergence of a rank four antisymmetric  \textit{bulk current} $K$
\begin{equation}
J^{\mu\nu\rho} (x)\equiv \partial_\lambda\, K^{\lambda\mu\nu\rho},
\label{bulk}
\end{equation}
where
\begin{equation}
K^{\lambda\mu\nu\rho}(x)\equiv \int_{V} \delta^{4)}\,\left[\,
x-z\,\right]\, dz^\lambda\wedge dz^\mu\wedge dz^\nu \wedge dz^\rho.
\end{equation} 
On the other hand,
\begin{equation}
  dz^\lambda\wedge dz^\mu\wedge dz^\nu \wedge dz^\rho=
  \epsilon^{\lambda\mu\nu\rho}\, d^4z\ ,
\end{equation}
so that one can write $ K^{\lambda\mu\nu\rho}(x)$ as
\begin{equation}
K^{\lambda\mu\nu\rho}(x)=\epsilon^{\lambda\mu\nu\rho}\,\Theta_{ V}(x),
\label{caratt}
\end{equation}
where
\begin{equation}
\Theta_{ V}(x)=\int_{V} d^4z \,\delta^{4)}\,\left[\,x-z\,\right],
\end{equation}
stands for the (generalized) unit step-function of the ${V}$ manifold, namely,
$\Theta_{ V}\left(\, P  \in V \,\right)=1\ ,\quad
\Theta_{ V}\left(\, P  \notin V \,\right)=0$.\\

By inverting Eq.(\ref{bulk}) one can also express the bulk-current $K$ in terms of the boundary current $J$
\begin{equation}
\partial_\lambda\, K_V^{\lambda\mu\nu\rho}=J_{\partial V}^{\mu\nu\rho}
\left(\,x\,\right)\longrightarrow
K_V^{\lambda\mu\nu\rho}=\partial^{[\,\lambda}\, \frac{1}{\partial^2}\,
 J_{\partial V}^{\mu\nu\rho\,]}, \label{bound}
\end{equation}
so that the general solution of Maxwell's field equation (\ref{maxx}) 
takes the following form
\begin{equation}
F^{\lambda\mu\nu\rho}= f\,\epsilon^{\lambda\mu\nu\rho}
+\kappa\,\partial^{[\,\lambda}\,
\frac{1}{\partial^2}\,J_{\partial V}^{\mu\nu\rho\,]}
= \epsilon^{\lambda\mu\nu\rho}\left(\, f  +
\kappa\, \Theta_V\left(\, x\,\right)\, \right), \label{class}
\end{equation}
where $f$ is, again, the constant solution of the homogeneous equation.

The above solution indicates that the interior and exterior regions of a "vacuum domain" are characterized by a different value of the vacuum energy density and pressure. This is the main feature of most phenomenological ``bag models'' of hadrons. The essential difference is that, in our case, this property is a dynamical consequence of the coupling of the 3-index potential to a relativistic test bubble, as dictated by the requirement of gauge invariance under the transformation (\ref{gauge}). In other words, \textit{the overall effect of the coupling is that the $A$-field is responsible for a long-range static interaction that results in the existence of a ground state with distinct phases.} It seems to us that this \textit{dynamical} property of the CSS-Lagrangian is a reminder of the topological structure of the ground state of gluodynamics. In other words, the many vacuums that arise in the topological sector of $QCD$ as a consequence of instanton effects reappear in the guise of "hadronic vacuum-domains," or glue bags, as a consequence of the vacuum polarization effects due to the CSS-potential.

\subsection{The CSS-potential satisfies the ``Wilson-loop'' criterion for confinement}
\label{W}

We have mentioned in the Introduction that YM-instantons alone do not generate a confining force between color charges. In contrast, one of the advantages of the field theory approach is that the CSS-Lagrangian has a built-in mechanism for confinement, i.e., it gives rise to a confining potential between infinitesimal surface elements of a gluebag. The crux of the argument is the calculation of the "Wilson loop" for the three-index potential coupled to the 
boundary current of a bag
\begin{eqnarray}
W\left[\, J\,\right]=\Big\langle\, \exp\left(-\frac{\kappa}{3!}\int d^4x
A_{\mu\nu\rho}\,
J^{\mu\nu\rho}\,\right)\, \Big\rangle=
\frac{Z\left[\, J\,\right]}{Z\left[\,0\,\right]}
\equiv \exp\left(-\Gamma\left[\, J\,\right]\,\right). \label{3.6b}
\end{eqnarray}

From the above expression it is possible to derive the expression of the static potential following the standard prescription

\begin{equation}
V\left(\, R \, \right)\equiv -\lim_{ T\to\infty  }\frac{1}{T}\ln \, 
W\left[\,\partial{B}\right]\ .\label{defv}
\end{equation}

In the case of a spherical bag, the calculations were carried out explicitly in Ref.\cite{Aurilia:2004fz}. 
For our present purposes, it is sufficient to recall the following general result: in order to extract the static 
potential $V\left(\, R\, \right)$ from Eq.(\ref{defv}) one must compute the contribution of the currents associated with a pair of antipodal points $P$ and $\overline P$ on the surface of the bag
\begin{eqnarray}
  \int_B d^4x\,
   J^{\mu\nu\rho} \, \frac{1}{\partial^2}\, J_{\,
   \mu\nu\rho}&&=
   \int_{\partial B} \int_{\partial B}
   dy^\mu\wedge dy^\nu\wedge dy^\rho\, \frac{1}{\partial^2}\,
   dy_\mu^\prime \wedge dy_\nu^\prime \wedge dy_\rho^\prime=\nonumber\\
&&=\frac{1}{4\pi^2}\,\int_0^T d\tau\int_T^0 d\tau^\prime
  \int_{ S^{(2)}  } d^2\sigma \int_{ S^{(2)}  } d^2\xi\,\times\nonumber\\
  && y^{\mu\nu\rho}\left(\,\tau\ ,\sigma \,\right)\, \frac{1}{\left[ \,
   y\left(\, \tau\ ,\sigma\,\right)- y\left( \, \tau^\prime\ ,\xi\,\right)
   \,\right]^2} \,
   y_{\mu\nu\rho}\left(\, \tau^\prime\ ,\xi \,\right)\,\delta^2\left[\,
   \xi -\sigma\,\right]\ .\nonumber\\
&& \label{double}
\end{eqnarray}

In the above expression, $\left(\, \sigma^1\ ,\sigma^2\,\right)$ and $\left(\, \xi^1\ ,\xi^2\,\right)$ are two independent sets of world coordinates on the boundary manifold. The world-history of these two points, in Euclidean time, constitutes the ``Wilson loop''. 

Close inspection of the double integral (\ref{double}) reveals that the interaction between two diametrically opposite surface elements is mediated by a Coulomb force that operates in the \textit{bulk} of the bag. This corresponds to the explicit form of the Green function in Eq. (\ref{double}). In the case of a spherical bubble, the radial dependence of the surface elements can be calculated explicitly and leads to the following result
\begin{equation}
V\left(\, R \, \right)\equiv -\lim_{ T\to\infty  }\frac{1}{T}\ln \, 
W\left[\,
\partial{B}\right]=\frac{\pi\, \kappa^2}{96}\, R^3\ .  \label{V(R)}
 \end{equation}

This expression represents an attractive potential that is proportional to the volume enclosed by the membrane. Again, it seems noteworthy that, unlike the phenomenological bag models where confinement is imposed at the outset, the result (\ref{V(R)})is the combined effect of bulk and boundary dynamics of the CSS-Lagrangian (\ref{med}) through the intermediary 3-index gauge potential.

\textit{The result (\ref{V(R)}) is also the basis of the noted correspondence with $QED_2$.} In order not to break the continuity of our discussion, the correspondence with the two-dimensional case will be discussed in more quantitative terms in Section IV. However, it may be helpful to anticipate here the basis of that correspondence: in Section IV, we argue that the zero-charge sector of $QED_2$ may also be interpreted as a \textit{bag model in $(1+1)$-dimensions.} Once this unconventional interpretation is clarified, it is almost straightforward to show that the same dynamics is effectively realized in $QCD$. Thus, we note that the ``volume law'' encoded in Eq.(\ref{V(R)}) represents a natural extension of the ``area law" for the Wilson loop of a quark-antiquark pair bounded by a string. In the string case, it is well known that the definition (\ref{defv}) leads to a linear potential between two test quarks
\begin{equation}
V(R)\propto\ R \ .\label{linear}
\end{equation}

The behavior of the Coulomb potential in $(1+1)$-dimensions meets this requirement. The linearly rising potential (\ref{linear}) is consistent with the ``string-picture '' of charge confinement, namely, that opposite charges are connected by a ``flux-tube'', or string, of constant energy per unit length. However, \textit{in one spatial dimension, strings and membranes reduce to a pair of points.} This geometric degeneracy is removed in $(3+1)$-dimensions. Evidently, it is the ``membrane-bag'' sector of $QED_2$ that is effectively realized in $QCD$ since the three-index potential is minimally coupled to closed membranes which, in turn, constitute the boundary of a bag.

We further observe that Eqs.(\ref{V(R)}) and (\ref{linear}) describe the same geometric effect. In both cases the static potential is proportional to the ``volume'' of the manifold connecting the two test charges. In Eq.(\ref{linear}), $R$ is essentially the "linear volume" enclosed by the two external sources. In our case, $R^3$ is proportional to the volume enclosed by the external membrane. In either case, the result reflects the basic underlying idea that "confinement", at the classical level, requires an infinite amount of energy in order to separate the two sources.

\section{Explicit computation of the static potential}

Inspired by the preceding observations, the purpose of this Section is to further elaborate on the physical content of the Chern-Simons term as a gauge invariant interaction associated with a Maxwell field of the fourth kind. To do this, we will work out the static potential for the $4$-D generalization of the Schwinger model, as originally introduced in Ref. \cite{Aurilia:1980jz}, via a path-integral approach. In effect, the initial point of our analysis is the bosonized form of the Schwinger model in D=$4$, that is,
\begin{equation}
{\cal L} = \frac{1}{2}{\left( {{\partial _\mu }\phi } \right)^2} + \frac{1}{2}m_\phi ^2{\phi ^2} + \frac{g}{{6\sqrt \pi  }}{\partial _\mu }\phi\ {\varepsilon ^{\mu \nu \rho \sigma }}{A_{\nu \rho \sigma }} - \frac{1}{{48}}F_{\mu \nu \rho \sigma }^2, \label{Scwhin3-05}
\end{equation}
where $g$ is a coupling constant and $m_\phi$ refers to the mass of the scalar field $\phi$.

According to usual procedure, integrating out the $\phi$ field induces an effective theory for the $A_{\nu \rho \sigma }$ field.  It is now important to recall that the $ A_{\nu \rho \sigma }$ field  can also be written as ${A_{\nu \rho \sigma }} = {\varepsilon _{\nu \rho \sigma \lambda }}{\partial ^\lambda }\xi$  \cite{Aurilia:2004cb, Aurilia:2004fz}, where $\xi$ refers to  another scalar field. This then leads to the following effective theory for the model under consideration:
\begin{equation}
{\cal L} = \frac{1}{2}\left[ {\xi \ \Delta \left( {1 + \frac{{{\raise0.7ex\hbox{${{g^2}}$} \!\mathord{\left/
 {\vphantom {{{g^2}} \pi }}\right.\kern-\nulldelimiterspace}
\!\lower0.7ex\hbox{$\pi $}}}}{{\left( {\Delta  - m_\phi ^2} \right)}}} \right)\Delta \ \xi } \right], \label{Scwhin3-10}
\end{equation}
where $\Delta  = {\partial _\mu }{\partial ^\mu }$.

We are now ready to compute the interaction energy between static pointlike sources. We start off our analysis by writing down the functional generator of the Green's functions, that is,
\begin{equation}
Z\left[ J \right] = \exp \left( { - \frac{i}{2}\int {{d^4}x{d^4}yJ(x)D(x,y)J(y)} } \right), \label{Scwhin3-15}
\end{equation}
where, $D(x,y) = \int {\frac{{{d^4}k}}{{{{\left( {2\pi } \right)}^4}}}D(k){e^{ - ikx}}}$, is the propagator. In this case, the corresponding propagator is given by
\begin{equation}
D(k) = \left( {1 - \frac{{m_\phi ^2}}{{{{\cal M}^2}}}} \right)\frac{1}{{{k^2}\left( {{k^2} + {{\cal M}^2}} \right)}} + \frac{{m_\phi ^2}}{{{{\cal M}^2}}}\frac{1}{{{k^4}}}, \label{Scwhin3-20}
\end{equation}
where ${{\cal M}^2} = m_\phi ^2 - {\raise0.5ex\hbox{$\scriptstyle {{g^2}}$}\kern-0.1em/\kern-0.15em
\lower0.25ex\hbox{$\scriptstyle \pi $}}$.

Enlisting the standard representation $Z = {e^{iW\left[ J \right]}}$ and employing Eq. (\ref{Scwhin3-15}), ${W\left[ J \right]}$ takes the form 
\begin{equation}
W[J] =  - \frac{1}{2}\int {\frac{{{d^4}k}}{{{{\left( {2\pi } \right)}^4}}}} J{\left( k \right)^ * }\left\{ \left( {1 - \frac{{m_\phi ^2}}{{{{\cal M}^2}}}} \right)\frac{1}{{{k^2}\left( {{k^2} + {{\cal M}^2}} \right)}} + \frac{{m_\phi ^2}}{{{{\cal M}^2}}}\frac{1}{{{k^4}}} \right\}J(k). \label{Scwhin3-25}
\end{equation}

Next, for $J({\bf x}) = \left[ {Q{\delta ^{\left( 3 \right)}}\left( {{\bf x} - {{\bf x}^{\left( 1 \right)}}} \right) + {Q^ \prime }{\delta ^{\left( 3 \right)}}\left( {{\bf x} - {{\bf x}^{\left( 2 \right)}}} \right)} \right]$, we obtain that the interaction energy of the system is given by
\begin{equation}
{V} =  - Q{Q^ \prime }\int {\frac{{{d^3}k}}{{{{\left( {2\pi } \right)}^3}}}} \left[ {\frac{{{\raise0.5ex\hbox{$\scriptstyle {{g^2}}$}
\kern-0.1em/\kern-0.15em
\lower0.25ex\hbox{$\scriptstyle \pi $}}}}{{{\raise0.5ex\hbox{$\scriptstyle {{g^2}}$}
\kern-0.1em/\kern-0.15em
\lower0.25ex\hbox{$\scriptstyle \pi $}} - m_\phi ^2}}\frac{1}{{{{\bf k}^2}\left( {{{\bf k}^2} + {\raise0.5ex\hbox{$\scriptstyle {{g^2}}$}
\kern-0.1em/\kern-0.15em
\lower0.25ex\hbox{$\scriptstyle \pi $}} - m_\phi ^2} \right)}} - \frac{{m_\phi ^2}}{{{\raise0.5ex\hbox{$\scriptstyle {{g^2}}$}
\kern-0.1em/\kern-0.15em
\lower0.25ex\hbox{$\scriptstyle \pi $}} - m_\phi ^2}}\frac{1}{{{{\bf k}^4}}}} \right]{e^{i{\bf k} \cdot {\bf r}}},   \label{Scwhin3-30}
\end{equation}
where ${\bf r} = {{\bf x}^{\left( 1 \right)}} - {{\bf x}^{\left( 2 \right)}}$.

This, together with ${Q^ \prime }=-Q$, yields finally
\begin{equation}
{V} = \frac{{{Q^2}}}{{4\pi }}\left[ {\frac{{{\raise0.5ex\hbox{$\scriptstyle {{g^2}}$}
\kern-0.1em/\kern-0.15em
\lower0.25ex\hbox{$\scriptstyle \pi $}}}}{{{{\left( {{\raise0.5ex\hbox{$\scriptstyle {{g^2}}$}
\kern-0.1em/\kern-0.15em
\lower0.25ex\hbox{$\scriptstyle \pi $}} - m_\phi ^2} \right)}^2}}}\frac{1}{L}\left( {1 - {e^{ - \sqrt {{\raise0.5ex\hbox{$\scriptstyle {{g^2}}$}
\kern-0.1em/\kern-0.15em
\lower0.25ex\hbox{$\scriptstyle \pi $}} - m_\phi ^2} \ L}}} \right) + \frac{{m_\phi ^2}}{{2\left( {{\raise0.5ex\hbox{$\scriptstyle {{g^2}}$}
\kern-0.1em/\kern-0.15em
\lower0.25ex\hbox{$\scriptstyle \pi $}} - m_\phi ^2} \right)}}L} \right], \label{Scwhin3-35}
\end{equation}
where $L = |{\bf r}|$. One immediately sees that the above static potential profile is analogous to that encountered in the two-dimensional Schwinger model. As a matter of fact, in order to put our discussion into the proper context, it is useful to summarize the relevant aspects of the two-dimensional Schwinger model. In such a case, we begin by recalling the bosonized form of the model under consideration \cite{Gross:1995bp}:
\begin{equation}
{\cal L} =  - \frac{1}{4}F_{\mu \nu }^2 + \frac{1}{2}{\left( {{\partial _\mu }\phi } \right)^2} - \frac{e}{{2\sqrt \pi  }}{\varepsilon ^{\mu \nu }}{F_{\mu \nu }}\phi  + m\sum \left( {\cos \left( {2\pi \phi  + \theta } \right) - 1} \right), \label{Scwhin3-40}
\end{equation}
where $\sum  = \left( {\frac{e}{{2{\pi ^{{\raise0.5ex\hbox{$\scriptstyle 3$}
\kern-0.1em/\kern-0.15em
\lower0.25ex\hbox{$\scriptstyle 2$}}}}}}} \right){e^{{\gamma _E}}}$ with ${\gamma _E}$ the Euler-Mascheroni constant and $\theta$ refers to the $\theta$-vacuum. 

Consequently, by using the gauge-invariant but path-dependent variables formalism which is known to provide a physically-based alternative to the Wilson loop approach \cite{Gaete:1999iy, Gaete:2001wh}, the static potential reduces to
\begin{equation}
V = \frac{{{Q^2}}}{2}\frac{{\sqrt \pi  }}{e}\left( {1 - {e^{ - \frac{e}{{\sqrt \pi  }}L}}} \right), \label{Scwhin3-45}
\end{equation}
for the massless case. On the other hand, for the massive case ($\theta=0$), the static potential then becomes
\begin{equation}
V = \frac{{{Q^2}}}{{2\lambda }}\left( {1 + \frac{{4\pi m\sum }}{{{\lambda ^2}}}} \right)\left( {1 - {e^{ - \lambda L}}} \right) + \frac{{{q^2}}}{2}\left( {1 - \frac{{{\raise0.5ex\hbox{$\scriptstyle {{e^2}}$}
\kern-0.1em/\kern-0.15em
\lower0.25ex\hbox{$\scriptstyle \pi $}}}}{{{\lambda ^2}}}} \right)L, \label{Scwhin3-50}
\end{equation}
where ${\lambda ^2} = \frac{{{e^2}}}{\pi } + 4\pi m\sum$. The above results clearly show that the $4$-D generalization of the Schwinger model is structurally identical to the $2$-D Schwinger model.

In this perspective it seems worth recalling that there is an alternative way of obtaining the Lagrangian density  (\ref{Scwhin3-10}), which provides a complementary insight into the physics of confinement. In fact, we refer to a theory of antisymmetric tensor fields that results from the condensation of topological defects as a consequence of the Julia-Toulouse mechanism. More precisely, the Julia-Toulouse mechanism is a condensation process dual to the Higgs mechanism proposed in \cite{Quevedo:1996uu}. This mechanism describes phenomenologically the electromagnetic behavior of antisymmetric tensors in the presence of magnetic-branes (topological defects) that eventually condensate due to thermal and quantum fluctuations. Using this phenomenology, we have discussed in \cite {Gaete:2004dn,Gaete:2005am} the dynamics of the extended charges (p-branes) inside the new vacuum provided by the condensate. Actually, in \cite {Gaete:2004dn} we have considered the topological defects coupled both longitudinally and transversally to two different tensor potentials, $A_p$ and $B_q$, such that $p+q+2=D$, where $D=d+1$ space-time dimensions. The technical details are given in Ref. \cite{Gaete:2004dn}. The net result is that, after the condensation, the Lagrangian density turns out to be
\begin{equation}
{\cal L} = \frac{{{{\left( { - 1} \right)}^q}}}{{2\left( {q + 1} \right)!}}{\left[ {{H_{q + 1}}\left( {{B_q}} \right)} \right]^2} + e{B_q}{\varepsilon ^{q,\alpha ,p + 1}}{\partial _\alpha }{\Lambda _{p + 1}} + \frac{{{{\left( { - 1} \right)}^{p + 1}}}}{{2\left( {p + 2} \right)!}}{\left[ {{F_{p + 2}}\left( {{\Lambda _{p + 1}}} \right)} \right]^2} + \frac{{{{\left( { - 1} \right)}^{p + 1}}\left( {p + 1} \right)!}}{2}{m^2}\Lambda _{p + 1}^2, \label{Scwhin3-55}
\end{equation}
showing a $B$$\wedge$$F$ type of coupling between the $B_q$ potential with the tensor $\Lambda_{p+1}$ carrying the degrees of freedom of the condensate. Following our earlier procedure \cite{Gaete:2004dn}, the effective theory that results from integrating out the fields representing the vacuum condensate, is given by
\begin{equation}
{\cal L} = \frac{{{{\left( { - 1} \right)}^{q + 1}}}}{{2\left( {q + 1} \right)!}}{H_{q + 1}}\left( {{B_q}} \right)\left( {1 + \frac{{{e^2}}}{{\Delta  - {m^2}}}} \right){H^{q + 1}}\left( {{B_q}} \right). \label{Scwhin3-60}
\end{equation}  
Hence we see that this expression with $p=-1$ and $q=3$ becomes
\begin{equation}
{\cal L} = \frac{1}{{2 \times 4!}}{F_{\mu \nu \rho \lambda }}\left( A \right)\left( {1 + \frac{{{e^2}}}{{\Delta  - {m^2}}}} \right){F^{\mu \nu \rho \lambda }}\left( A \right). \label{Scwhin3-65}
\end{equation}
It is straightforward to verify that Eq. (\ref{Scwhin3-65}) reduces to Eq. (\ref{Scwhin3-10}).

In this way we establish a new connection among different effective theories. From this discussion it should be clear that the above connections are of interest from the point of view of providing unifications among diverse models as well as exploiting the equivalence in explicit calculations.

\section{Final Remarks}

The Schwinger model of two-dimensional "electrodynamics", because of 
its \textit{kinematical} constraints, has a built-in mechanism of 
confinement that has inspired many subsequent models of 
"hadronization" since the very inception of $QCD$ as the leading 
theory that describes the interaction among colored quarks and 
gluons.

The possibility of generalizing $QED_2$ to four dimensions has always 
been met with skepticism, mostly because the "obvious" generalization 
of the model has been known for a long time, namely, the theory of 
ordinary quantum electrodynamics, or $QED_4$. As a matter of 
historical fact, the "Schwinger model'' was conceived by restricting 
the familiar Maxwell-Dirac Lagrangian to $(1+1)$-dimensions with an 
eye on the relationship between gauge invariance and mass. This 
original procedure of "descending'' from four to two dimensions, 
while keeping the same form of the Maxwell-Dirac Lagrangian, has 
generated the widespread notion that $QED_4$ is the \textit{unique} 
extension of $QED_2$. It has been recognized, however, that this is 
not a one-to-one correspondence so that, by the reverse procedure of 
"ascending'' from two to four dimensions, the ``electrodynamic'' 
interpretation of the Schwinger model seems purely formal in the 
sense that \textit{the physical content} of $QED_2$ and $QED_4$ is 
drastically different. Thus, in order to underscore this difference, 
in Section IV we have itemized the essential \textit{physical} 
properties of the Schwinger model. Those properties, we have argued, 
pertain to a \textit{theory of spatially extended objects, membranes 
and bags to be precise, and can be realized in two as well as in four 
dimensions}. It is a remarkable fact, and the main conclusion of this 
paper, that the very dynamics of Quantum Chromodynamics in the large 
distance limit possesses the same kinematical constraints, and 
therefore the same confining and screening potential that are found 
in the two-dimensional case. This result hinges on a 
reinterpretation of the Chern-Simons term in $QCD$ as a gauge 
invariant interaction associated with a Maxwell field of the fourth 
kind. The gauge invariant coupling of the CSS-potential requires the 
existence of relativistic closed membranes.  We have treated such 
membranes as an approximation to the physical boundary of non 
perturbative gluebags. The dynamics of this system has been shown to 
be exactly solvable. The static interaction potential, unlike the 
instanton configurations of the old "topological'' interpretation, 
satisfies the Wilson-loop criterion for confinement while the quantum 
axial anomaly in $QCD$ triggers the same screening mechanism of mass 
generation that operates in the two-dimensional Schwinger model.

\section{acknowledgements}

P. Gaete was partially supported by FONDECYT (Chile) grant 1130426, DGIP (UTFSM) internal project USM 111458. One of us (PG) wants to thank the Abdus Salam ICTP for hospitality and support. This work was carried out in part at the Physics Department, Harvey Mudd College, and in part at the Physics Department of the University of Trieste. One of us (AA) wishes to thank John Townsend (HMC) and Euro Spallucci (U of T) for the hospitality extended to him while on sabbatical leave from California State Polytechnic University-Pomona.

\section{Appendix}

\subsection{Quantum anomaly and mass generation: The massive case}
\label{massive}

The purpose of this Appendix is to highlight some other aspects of the strict correspondence between the two-dimensional case and the four-dimensional one. Two interesting questions arise naturally. 
First, what happens if the CSS-potential acquires a 
mass? And, concomitantly, what is the physical mechanism that may 
induce a mass term for $A_{\mu\nu\rho}$?

Summarizing our previous and subsequent discussion, the following 
properties of $A_{\mu\nu\rho}$ constitute the crux of the 
\textit{generalized Schwinger mechanism in $QCD$}: \\
\\
a) When massless, $A_{\mu\nu\rho}$ represents nothing more than a 
constant background field since, as we have shown, in 
(3+1)--dimensions $A_{\mu\nu\rho}$ does not possess radiative degrees 
of freedom. Here, for later reference, we wish to add that this 
property, even though peculiar, is shared by all $d$--potential forms 
in $(d+1)$--spacetime dimensions \cite{Aurilia:1978qs,Aurilia:1980xj}.\\
For instance, the best known case is in two dimensions:
$F_{\mu\nu}=\partial_{[\,\mu}A_{\nu]}=\epsilon_{\mu\nu}\, f$ while in
four dimensions, $F_{\mu\nu\rho\sigma}=\epsilon_{\mu\nu\rho\sigma}\, f$,
and $f$ represents a constant vacuum energy density in both cases by 
virtue of the field equations.\\
\\
b) If the field acquires a mass, then it describes \textit{massive 
pseudoscalar particles,} in two a well as in four dimensions. The two 
dimensional case is well known from the Schwinger model; in four 
dimensions, the free field equation for $A_{\mu\nu\rho}$ in the 
massive case \cite{Aurilia:1981xg,Aurilia:2004fz}

\begin{eqnarray}
	&&\partial^\lambda \partial_{\,[\,\lambda}
	A_{\mu\nu\rho\,]}+m^2
         A_{\mu\nu\rho}=0\ , \quad \Longrightarrow \partial^\mu \,
         A_{\mu\nu\rho}=0,
         \label{unobis}
         \end{eqnarray}
imposes the divergence free constraint on the four components of 
$A_{\mu\nu\rho}$ leaving only one {\it propagating} degree of 
freedom. \textit{In other words, the introduction of a mass term 
``excites'' a dynamical (pseudoscalar) particle of matter out of the 
constant energy background.} \footnote{An interesting cosmological 
application of this transmutation was discussed in Ref.(\cite{Ansoldi:2001xi}) 
in  connection with the problem of ``dark matter/energy in the universe. 
There, we also discussed the subtle issue of the presence of a mass 
term in an otherwise gauge invariant theory through the use of 
Stueckelberg's formalism.}\\
\\
c) Evidently, the transition from case a) (massless, non dynamical field)
to case b) (massive particles) requires a physical mechanism for its 
enactment. In the case at hand, we will show that the mass-inducing 
term for the CSS-potential is the \textit{quantum anomaly term} in 
$QCD$ in perfect analogy with the two dimensional case of the 
Schwinger model. In the two dimensional case the occurrence of mass 
appears in the form of a pole in the propagator of the gauge field; 
in $QCD$, it manifests itself as the ``Kogut-Susskind'' pole in the 
correlation function of the topological charge density \cite{Kogut:1974ni}.

\subsection{From two to four dimensions}

At the quantum level, the result of the previous section is altered 
by the phenomenon of ``screening.'' Quantum mechanically, rather than 
supplying the large amount of energy that is required in order to 
separate two test charges, it is energetically more favorable to 
create quark-antiquark pairs within the volume separating the two 
test charges. This phenomenon was anticipated long ago by Kogut and 
Susskind \cite{Kogut:1974sn}, again in connection with the Schwinger model 
of $QED_2$. The net physical result of the quantum anomaly in $QED_2$, is the 
``hadronization process,'' namely the binding of quarks into physical 
hadrons through screening of the original test charges and 
\textit{the emergence of mass, corresponding to a pole in the 
propagator of the gauge field} .
 
A qualitative argument that 
anticipates the existence of a similar mechanism in the large 
distance limit of $QCD$ is based on the existence of a pole, the so 
called Kogut-Susskind pole, in the correlation function of the 
topological charge density $Q$:
\begin{equation}
\chi =i\lim_{q\rightarrow 0}\, q^\mu\, q^\nu \,\int d^4x \, e^{iqx}\,
\langle 0\, \vert\, T\left[\,
K_\mu(x)\, K_\nu(0)\,\right]\, \vert\, 0\rangle.  \label{suscep}
\end{equation}

In Gabadadze's work, this correlation function is referred to as 
\textit{vacuum topological susceptibility}. The expression of $\chi$ 
quoted above is derived by recalling that the topological charge 
density $Q$ is the derivative of the Chern-Simons current 
$Q=\partial_\mu\, K^\mu$ and substituting this identity into the 
expression for the correlation function of the topological charge 
density.

We have noticed in subsection \ref{massive} that, in agreement with Luscher's 
observation, the correlation function of two boundary current is non 
zero and corresponds to the existence of a Coulomb-type potential 
within the bulk of the hadronic vacuum domain $V$. This 
implies that the above expression for $\chi$ is non zero and this can 
only happen if the correlation function
of two Chern-Simons currents develops a pole in the limit of 
vanishing momentum.
This is one way in which the ``Schwinger mechanism'' of mass 
generation is implemented in an otherwise gauge invariant theory.\\

In this Subsection, following the example of $QED_2$ which we shall
discuss in Section IV.D, we wish to show that the \textit{screening} of the 
classical potential (\ref{V(R)}), extracted above from the Wilson 
loop, takes place when the effect of the quantum axial current 
anomaly in $QCD$ is taken into account.

For the sake of comparing our results, it seems useful to recall, at 
this point, the main properties that hold in $2D$ as we discuss them 
in the ``bag interpretation'' of $QED_2$.

In the presence of fermions, it is well known that the vector and 
axial-vector currents cannot be simultaneously conserved. If one 
insists on the conservation of the vector current, then the 
divergence of the axial-vector current is not zero. Without 
digressing on these well known results, the properties of 
bubble-dynamics in $2D$ may be distilled into the following points:\\

\begin{enumerate}

\item
There is a vector current, $j^\mu(x)$, coupled to a vector gauge 
potential $A_\mu(x)$.

\item
The Maxwell field $F_{\mu\nu}=\partial_{[\,\mu}\, A_{\nu\,]}$ , in 
this case \textit{a differential two-form in two spacetime 
dimensions,} is dual to a zero-form:
$F_{\mu\nu}=  \epsilon_{\mu\nu}\, f$.\\
In the absence of coupling, $f$ amounts to an arbitrary integration 
constant that is physically related to a background vacuum energy 
density. When coupled to a test bubble, or, in electrodynamics 
parlance, to a dipole charge distribution, the gauge field $A$ gives 
rise to a confining potential that is proportional to the linear 
volume of the bag.

\item
The axial current is dual to the vector current: $j^{5\,\mu}(x)\equiv 
\epsilon^{\mu\nu}\, j_\nu(x)$.

\item
The axial current $j^{5\,\mu}(x)$ is ``anomalous'', i.e., its 
divergence is proportional to the dual field strength of $A_\mu(x)$, 
that is, $\partial_\mu\, j^{5\,\mu}(x)\propto 
\epsilon^{\nu\rho}\,\partial_{[\,\nu} \, A_{\rho\,]}\propto
\epsilon^{\nu\rho}F_{\nu\rho} $.

\end{enumerate}

The purpose of this brief synopsis of the salient features of $QED_2$ 
is to motivate the calculations described below for the 
CSS-Lagrangian. Indeed, a strong argument for the viability of the 
Schwinger mechanism in $QCD$ is that the three-index gauge potential 
$A_{\mu\nu\rho}$ possesses the same \textit{physical properties} as 
the gauge vector potential $A_\mu$ in two dimensions. Thus, in order 
to underscore the stringent similarity between the two cases, we list 
below the corresponding items that we wish to implement in $QCD$:

\begin{enumerate}

\item
There is a three-index \textit{tensor} current, 
$J^{\mu\nu\rho}\left(\, x\,\right)$, 
coupled to a gauge three-form potential $A_{\mu \nu\rho }(x)$.

\item
The Maxwell field $F_{\mu\nu\rho\sigma}=\partial_{[\,\mu }\,
A_{\nu\rho\sigma\,]}$ in this case \textit{a differential four-form 
in four dimensions,} is dual to a zero-form:
$F_{\mu\nu\rho\sigma}=\partial_{[\,\mu}\, A_{\nu\rho\sigma\,]}=
\epsilon_{\mu\nu\rho\sigma}f$.\\
In the absence of coupling, as we have seen, $f$ amounts to an 
arbitrary integration constant that is physically related to a 
background energy density. When coupled to a test bubble, the gauge 
field $A$ gives rise to a confining potential that is proportional to 
the volume of the bag.

\item
The role of ``axial current'' is played by the Hodge dual of the 
tensor current, namely, 
\begin{equation}
j^{5\,\mu}(x)\equiv
\epsilon^{\mu\nu\rho\sigma}\, j_{\nu\rho\sigma}(x)
\label{jjdual}.
\end{equation}
\item
The axial current $j^{5\,\mu}(x)$ is ``anomalous'', i.e., its divergence is
proportional to the dual field strength of $A_{\lambda  \mu \nu }(x)$, that is,
$\partial_\mu\, j^{5\,\mu}(x)\propto
\epsilon^{\mu\nu\rho\sigma }\,
\partial_{[\,\lambda }\, A_{\nu\,\rho\,\sigma\,]}\propto
\epsilon^{\lambda\mu\nu\rho} F_{\lambda\mu\nu\rho} $.

\end{enumerate}

\subsection{Screening and mass generation in $QCD$}

Our immediate objective is to translate the above correspondence into 
explicit computational steps and show that the same screening 
mechanism that generates mass in two dimensions is also active in the 
case of the CSS-Lagrangian.
Thus, according to the above code of correspondence, one needs two 
currents that are dual to each other and an expression of the quantum 
anomaly in terms of the field strength of the abelian CSS-potential. 
It is a remarkable fact that these necessary ingredients emerge 
naturally in the field theoretical interpretation of the topological 
charge density. Indeed, we recall from Section II that the 
$\theta$-term in the YM-action can be rewritten as an interaction term between 
the CSS-potential and the boundary current (\ref{sint}).

However, that expression is equivalent to
\begin{equation}
S_\theta=\frac{\kappa}{3!}\int d^4x\,
J^{\mu\nu\rho}_{\partial V}(x)\, A_{\mu\nu\rho}\ ,\qquad \kappa\equiv
\frac{3\theta}{4\pi^2}\, \Lambda_{QCD}^2,
\end{equation}
and the membrane current, being the boundary current of a bag, is
divergence-free

\begin{equation}
\partial_\mu\, J^{\mu\nu\rho}_{\partial V}=0.
\end{equation}

On the other hand, from $QCD$ we know that the \textit{color singlet} 
axial current is anomalous
\begin{eqnarray}
\partial_\mu\,  J^{5\, \mu} && =
\frac{g^2}{16\pi^2}\, \epsilon^{\lambda\mu\nu\rho}\,
\mathrm{Tr}\left(\,
\mathbf{F}_{\lambda\mu}\,\mathbf{F}_{\nu\rho}\,\right)\label{35},\\
F^a{}_{\lambda\mu} &&=\partial_{[\, \lambda}\, A^a{}_{ \mu\,]}
+g\, f^a{}_{bc}\, \left[\,   A^b{}_{ \lambda}\ ,  A^c{}_{ \mu} \,\right].
\label{4danom}
\end{eqnarray}

Therefore, in terms of the CSS-potential, the anomaly equation takes the form
\begin{equation}
\partial_\mu
J^{5\mu}=\frac{g^2\Lambda^2_{QCD}}{16\pi^2}\, \epsilon^{\lambda\mu\nu\rho}\,
\partial_{[\, \lambda}\, A_{\mu\nu\rho\,]}\ .\nonumber\\
\end{equation}

The net result of the above manipulations is that we reproduce the 
pattern of dual currents that exists in the $2$-dimensional case. 
Then, in exact analogy with the case of $QED_2$, 
we proceed to encode the constraints (\ref{jjdual}), (\ref{4danom}) 
in the generating functional as follows \footnote{From a mathematical 
point of view, it is important to realize that, once the expression 
of the anomaly is given, one can by-pass the integration over the 
fermionic degrees of freedom in the path-integral so long as one 
takes into account the constraints listed above. More specifically, 
the summation in the path integral must take into account the duality 
relation between $J^5$ and $J$ and the anomalous divergence as 
\textit{constraints} on the integration measure $[DJ^5]$. In other 
words, we take stock of the fact that, in the large distance limit of 
$QCD$, as well as in $QED_2$, the only remnant of quark-dynamics is 
encoded in the given anomaly equation.}:
\begin{equation}
\int D\left[\, A\, \right]\exp\left(-iS_0\left[A\right]-i\,
S_\theta\left[\, A\ ,
J_{\partial V}\,\right]\,\right)\equiv
Z\left[\, J\,\right],
\end{equation}

\begin{eqnarray}
Z &&=\int [DA][DJ^5]
\, \delta\left[\, J^5{}_\lambda-\frac{\Lambda^2_{QCD} }{3!}\,
\epsilon_{\lambda\mu\nu\rho}
\, J^{\mu\nu\rho}_{\partial V}\,\right]\, \delta\left[\, \partial_\mu\,
J^5{}^\mu -
\frac{g^2}{16\pi^2}\, \epsilon^{\lambda\mu\nu\rho}\,
Tr\left(\, F_{\lambda\mu}\,F_{\nu\rho}\,\right)\,\right]\times\nonumber\\
&&\exp\left(\, i S_0[A] +i \frac{3\theta}{4\pi^2}\frac{1}{3!}\int d^4x
J^{\lambda\mu\nu}_{\partial V}\mathrm{Tr}
\left(\, \mathbf{A}_{[\,\lambda}\mathbf{D}_\mu\, \mathbf{A}_{\nu\,]}\,\right)
\,\right).
\end{eqnarray}
Integration over $J^{5\,\mu}$ implements the anomaly relation
\begin{equation}
J^{\mu\nu\rho}_{\partial V}\equiv
\frac{g^2}{16\pi^2\Lambda^2_{QCD}}\epsilon^{\mu\nu\rho\sigma}\,
\partial_\sigma\frac{1}{\partial^2}
\mathrm{Tr}\left(\, \mathbf{F}_{\alpha\beta}{}^\ast
\mathbf{F}^{\alpha\beta}\,\right),
\end{equation}
while an integration by parts leads to the following expression for 
the generating functional
\begin{eqnarray}
Z &&=\int [DA]
 \exp\left(\,iS_0[A] + i\,\frac{\theta g^2}{128\pi^4\Lambda^2_{QCD}} \int
d^4x\left[\, \epsilon^{\lambda\mu\nu\rho}
\mathrm{Tr}\, \partial_{[\,\lambda} \left(\, \mathbf{A}_\mu\mathbf{D}_\nu\,
\mathbf{A}_{\rho\,]}\,\right)
  \frac{1}{-\partial^2}
\mathrm{Tr}\left(\, \mathbf{F}_{\beta\gamma}{}^\ast
\mathbf{F}^{\beta\gamma}\,\right)
\,\right] \,\right).\nonumber\\
&&
\end{eqnarray}

At large distance, or in the infrared limit, the second term in the 
above expression dominates over the Yang-Mills term when
\begin{equation}
\frac{\theta\, g^2}{128 \pi^4}\frac{\,\Lambda^2_{QCD} }{\vert k^2\vert}> 
\frac{1}{48 g^2},\label{ir}
\end{equation}
Equation (\ref{ir}) defines the distance where the anomaly induced action
dominates over the kinetic Yang-Mills term, i.e., it defines the infrared
range of momenta for which the topological term is dominant
\begin{eqnarray}
&& Z\approx\int [DA]
\exp\left(\, i\,\frac{\theta g^2}{128\pi^4\Lambda^2_{QCD}} \int
d^4x\left[\, \epsilon^{\lambda\mu\nu\rho}\partial_\lambda\,\mathrm{Tr}
\left(\, \mathbf{A}_{[\,\mu}\mathbf{D}_\nu\, \mathbf{A}_{\rho\,]}\,\right)
  \frac{1}{-\partial^2}
\mathrm{Tr}\left(\, \mathbf{F}_{\beta\gamma}{}^\ast
\mathbf{F}^{\beta\gamma}\,\right)
\,\right] \,\right).\nonumber\\
&&
\end{eqnarray}

Thus, in the infrared domain the Yang-Mills field enters the 
generating functional only through the abelian CSS-potential
\begin{eqnarray}
Z &&=\int [DA_\mu] \, [ DA_{\rho\sigma\tau}]
\delta\left[\, A_{\mu\nu\rho}  -\frac{1}{\Lambda_{QCD}^2}\,
\left(\,A^a_{[\, \mu}\, \partial_\nu\,
A^a_{\rho\,]} +\frac{2g}{3}\, f^{abc}\, A^a_{[\, \mu}\, A^b_\nu\,
A^c_{\rho \, ]}\,\right)\,\right]\times\nonumber\\
&&
\exp\left(\, i\,\frac{\theta g^2}{128\pi^4\Lambda^2_{QCD}} \int
d^4x\left[\, \epsilon^{\lambda\mu\nu\rho}\partial_{[\,\lambda}\,\mathrm{Tr}
\left(\, \mathbf{A}_\mu \mathbf{D}_\nu\, \mathbf{A}_{\rho\,]}\,\right)
  \frac{1}{-\partial^2}
\mathrm{Tr}\left(\, \mathbf{F}_{\beta\gamma}{}^\ast
\mathbf{F}^{\beta\gamma}\,\right)
\,\right] \,\right)\nonumber\\
&&=\int D[ A_{\rho\sigma\tau}]\exp\left(\, i \frac{1}{2\times 4!}
\int d^4x \, \partial_{[\,\lambda}\, A_{\mu\nu\rho\,]}\,
\frac{12g^2\kappa}{-\partial^2} \, \partial^{[\lambda}\,
A^{\mu\nu\rho\,]}\,\right).\nonumber\\
&&
\end{eqnarray}
Note that the Hodge dual of $J^{5\, \mu}$ is a rank-three, totally 
anti-symmetric current $J^{\mu\nu\rho}_{\partial V}$ so that in the 
non-perturbative  (strong-coupling) regime of $QCD$ we recover the 
same duality relationship that holds true in $QED_2$

\begin{equation}
 J^5{}_\mu = \frac{2g^2\kappa}{3!}
\epsilon_{\mu\nu\rho\lambda}\, J_{\partial V}^{\nu\rho\lambda}.
  \end{equation}

With the above results in hand, we conclude that the complete 
effective action in the large distance limit of $QCD$, 
\textit{including the effects of the quantum anomaly,} is as follows
\begin{equation} Z=\int [DA...]\, \exp\left(\, i\,\int d^4x\left[\,
-\frac{1}{2\times 4!} \partial_{[\,\lambda } A_{\mu\nu\rho\,]} \left(\,1 +
\frac{12g^2\kappa}{-\partial^2}\,\right)
  \partial^{[\,\lambda }\, A^{\mu\nu\rho\,]}\,\right]\,\right).
\label{effe4}
\end{equation}
The above expression represents the effective gauge invariant 
action for a \textit{massive} three-index potential anticipated in 
subsection \ref{W}. The physical spectrum consists of massive 
pseudo-scalar particles in exact analogy to the $2$-dimensional case. 
This is the Schwinger mechanism that operates in the strong coupling 
limit of $QCD$. 

\subsection{On the analogy between the ``topological'' terms in $2D$ 
and in $4D$.}

The ``topological term'' in two dimensions
\begin{equation}
S_\theta = \frac{\theta e}{4\pi}\int_V d^2x
\, \epsilon_{\mu\nu}\, F^{\mu\nu},
\end{equation}
is also given by a total divergence
\begin{equation}
S_\theta=\frac{\theta e}{2\pi}\int_V
d^2x \, \partial_\mu \, A^{5\mu},
\end{equation}
where $A^{5\mu}\equiv\epsilon^{\mu\nu}\, A_\nu $. From Stokes theorem it
follows that
\begin{equation}
S_\theta =\frac{\theta e}{2\pi}\oint_{\partial V} dy^\mu A_\mu(y)
=\frac{\theta e}{2\pi}\int_0^l ds\frac{ dy^\mu}{ds} A_\mu(y)\ ,\qquad
y^\mu(l)=y^\mu(0).
\end{equation}

The above equation can be re-written as
\begin{equation}
S_\theta=\frac{\theta e}{2\pi}\int d^2x\, J^\mu_{\partial V} A_\mu,
\end{equation}
where we have introduced the \textit{boundary current}
\begin{equation}
J^\mu_{\partial V}(x)=\oint_{\partial V}dy^\mu \,\delta^{(2)}(\, x-y\,).
\end{equation}

In $2D$ the topological term is equivalent to the coupling between
$A_\mu$ and the boundary current $J^\mu_{\partial V}$.  At large 
distance, or in the infrared limit, it follows from the effective action
for $QED_2$ that the scale at which the topological terms dominates
over the Maxwell term is given by
\begin{equation}
\frac{e^2}{\vert\, k^2\,\vert} >> 1 \longrightarrow \frac{1}{\vert k \vert}
>>  \frac{1}{e}.
\end{equation}

In this large distance regime we have that
\begin{equation}
J^\mu_{\partial V} =\epsilon^{\mu\nu}\, {}^\ast J_\nu= \frac{\pi}{\theta}
J^{5\,\mu},
\end{equation}
and ${}^\ast J^\nu$ is directed along the normal to $\partial V$,
\begin{equation}
\partial_\mu {}^\ast J^\mu=\partial_\mu \langle\, 0\,\vert\,
J^{5\mu}\,\vert\, 0\,
\rangle=\frac{e}{2\pi}\epsilon^{\alpha\beta}\, F_{\alpha\beta}.
\end{equation}
This allows us to express the v.e.v. of the vector current in terms of the
field strength
\begin{equation}
J^\mu_{\partial V}
=\frac{e}{2\pi}\epsilon^{\mu\nu}\partial_\nu\,\frac{1}{\partial^2}\,
\epsilon^{\alpha\beta}F_{\alpha\beta}.
\end{equation}
Therefore, the topological action becomes
\begin{equation}
S_\theta=-\frac{e^2}{4\pi}\int d^2x
F_{\mu\nu}\frac{1}{-\partial^2}F^{\mu\nu}.
\end{equation}

Finally, adding a gauge invariant kinetic term for the field $A_\mu$
gives us the total action
\begin{equation}
S_{tot}\equiv S_0 + S_\theta =-\frac{1}{4}\int d^2x F_{\mu\nu}\frac{
-\partial^2+  e^2/\pi}{-\partial^2} F^{\mu\nu}.
\end{equation}


\begin{thebibliography}{99}

\bibitem{Hasenfratz:1977dt}
  P.~Hasenfratz and J.~Kuti,
  Phys.\ Rept.\  {\bf 40}, 75 (1978).

\bibitem{Gabadadze:1997kj}
  G.~Gabadadze,
  Phys.\ Rev.\  D {\bf 58}, 094015 (1998).

\bibitem{Luscher:1978rn}
  M.~Luscher,
  Phys.\ Lett.\  B {\bf 78}, 465 (1978).


\bibitem{Jackiw:1976pf}
  R.~Jackiw and C.~Rebbi,
  Phys.\ Rev.\ Lett.\  {\bf 37}, 172 (1976).

\bibitem{Callan:1976je}
  C.~G.~.~Callan, R.~F.~Dashen and D.~J.~Gross,
  Phys.\ Lett.\  B {\bf 63}, 334 (1976).

\bibitem{Belavin:1975fg}
  A.~A.~Belavin, A.~M.~Polyakov, A.~S.~Schwartz and Yu.~S.~Tyupkin,
  Phys.\ Lett.\  B {\bf 59}, 85 (1975).

\bibitem{Wilson:1974sk}
  K.~G.~Wilson,
  Phys.\ Rev.\  D {\bf 10}, 2445 (1974).

\bibitem{Aurilia:1979dw}
  A.~Aurilia,
  Phys.\ Lett.\  B {\bf 81}, 203 (1979).


\bibitem{Maldacena:1997re}
  J.~M.~Maldacena,
  Adv.\ Theor.\ Math.\ Phys.\  {\bf 2}, 231 (1998).
  

\bibitem{Aharony:1999ti}
  O.~Aharony, S.~S.~Gubser, J.~M.~Maldacena, H.~Ooguri and Y.~Oz,
  Phys.\ Rept.\  {\bf 323}, 183 (2000).

\bibitem{Witten:1998zw}
  E.~Witten,
  Adv.\ Theor.\ Math.\ Phys.\  {\bf 2}, 505 (1998).

\bibitem{Gabadadze:1997zc}
  G.~Gabadadze,
  Phys.\ Rev.\  D {\bf 58}, 055003 (1998).


\bibitem{Schwinger:1962tn}
  J.~S.~Schwinger,
  Phys.\ Rev.\  {\bf 125}, 397 (1962).

\bibitem{Schwinger:1962tp}
  J.~S.~Schwinger,
  Phys.\ Rev.\  {\bf 128}, 2425 (1962).

\bibitem{Casher:1973uf}
  A.~Casher, J.~B.~Kogut and L.~Susskind,
  Phys.\ Rev.\ Lett.\  {\bf 31}, 792 (1973).



\bibitem{Coleman:1975pw}
  S.~R.~Coleman, R.~Jackiw and L.~Susskind,
  Annals Phys.\  {\bf 93}, 267 (1975).
  
  
  \bibitem{Aurilia:1977jz}
  A.~Aurilia and F.~Legovini,
  Phys.\ Lett.\  B {\bf 67}, 299 (1977).

  
\bibitem{Aurilia:1978qs}
  A.~Aurilia, D.~Christodoulou and F.~Legovini,
  Phys.\ Lett.\  B {\bf 73}, 429 (1978).

\bibitem{Aurilia:1979dx}
A.~Aurilia and D.~Christodoulou,
  J.\ Math.\ Phys.\  {\bf 20}, 1446 (1979).



\bibitem{Aurilia:1978yc}
  A.~Aurilia and D.~Christodoulou,
  Phys.\ Lett.\  B {\bf 78}, 589 (1978).




\bibitem{Aurilia:1980jz}
  A.~Aurilia, Y.~Takahashi and P.~K.~Townsend,
  Phys.\ Lett.\  B {\bf 95}, 265 (1980).
  

\bibitem{Aurilia:1980xj}
  A.~Aurilia, H.~Nicolai and P.~K.~Townsend,
  Nucl.\ Phys.\  B {\bf 176}, 509 (1980).

\bibitem{Aurilia:1983ih}
  A.~Aurilia, G.~Denardo, F.~Legovini and E.~Spallucci,
  Phys.\ Lett.\  B {\bf 147}, 258 (1984).

\bibitem{Aurilia:1984cm}
  A.~Aurilia, G.~Denardo, F.~Legovini and E.~Spallucci,
  Nucl.\ Phys.\  B {\bf 252}, 523 (1985).

\bibitem{Aurilia:1987cp}
  A.~Aurilia, R.~S.~Kissack, R.~B.~Mann and E.~Spallucci,
  Phys.\ Rev.\  D {\bf 35}, 2961 (1987).

\bibitem{Aurilia:1989sb}
  A.~Aurilia, M.~Palmer and E.~Spallucci,
  Phys.\ Rev.\  D {\bf 40}, 2511 (1989).

\bibitem{Ansoldi:1997hz}
  S.~Ansoldi, A.~Aurilia, R.~Balbinot and E.~Spallucci,
  Class.\ Quant.\ Grav.\  {\bf 14}, 2727 (1997).


\bibitem{Ansoldi:2000ge}
  S.~Ansoldi, C.~Castro and E.~Spallucci,
  Phys.\ Lett.\  B {\bf 504}, 174 (2001).


\bibitem{Ansoldi:2001xi}
  S.~Ansoldi, A.~Aurilia and E.~Spallucci,
  Phys.\ Rev.\  D {\bf 64}, 025008 (2001).

\bibitem{Aurilia:1981xg}
  A.~Aurilia and Y.~Takahashi,
  Prog.\ Theor.\ Phys.\  {\bf 66}, 693 (1981).

 
\bibitem{Aurilia:2004cb}
  A.~Aurilia and E.~Spallucci,
  Phys.\ Rev.\  D {\bf 69}, 105004 (2004).
  
  
\bibitem{Aurilia:2004fz}
  A.~Aurilia and E.~Spallucci,
  Phys.\ Rev.\  D {\bf 69}, 105005 (2004).
 
 
\bibitem{Kogut:1974ni}
  J.~B.~Kogut and L.~Susskind,
  Phys.\ Rev.\  D {\bf 9}, 697 (1974).

\bibitem{Kogut:1974sn}
  J.~B.~Kogut and L.~Susskind,
  Phys.\ Rev.\  D {\bf 9}, 3501 (1974).


\bibitem{Fujikawa:1983bg}
  K.~Fujikawa,
  Phys.\ Rev.\  D {\bf 29}, 285 (1984).

\bibitem{Bousso:2000xa}
  R.~Bousso and J.~Polchinski,
  JHEP {\bf 0006}, 006 (2000).
  
\bibitem{Dvali:2004tma}
  G.~Dvali,
  Phys.\ Rev.\  D {\bf 74}, 025018 (2006).
  
\bibitem{Gross:1995bp} D.~J.~Gross, I.~R.~Klebanov, A.~V.~Matytsin and A.~V.~Smilga,
Nucl.\ Phys.\ B {\bf 461}, 109 (1996).    
  
\bibitem{Gaete:1999iy} P.~Gaete and I.~Schmidt,
Phys.\ Rev.\ D {\bf 61}, 125002 (2000).

\bibitem{Gaete:2001wh} P.~Gaete and I.~Schmidt,
Phys.\ Rev.\ D {\bf 64}, 027702 (2001).  
  
\bibitem{Quevedo:1996uu} F. Quevedo and C. A. Trugenberger,
Nucl.Phys. B {\bf 501}, 143 (1997).

\bibitem{Gaete:2004dn} P. Gaete and C. Wotzasek, 
Phys.Lett. B {\bf 601}, 108, (2004).

\bibitem{Gaete:2005am}  P. Gaete and C. Wotzasek, 
Phys.Lett. B {\bf 634}, 545, (2006).  
  




\end{thebibliography}
\end{document}